\documentclass[twocolumn,showpacs,preprintnumbers,aps,amsmath,amssymb]{revtex4}
\usepackage{graphicx}
\usepackage{dcolumn}
\usepackage{amsmath}
\def\apj #1 #2 #3 {#1, ApJ, {\bf #2}, #3}
\def\apjl #1 #2 #3 {#1, ApJ, {\bf #2}, L#3}
\def\apjs #1 #2 #3 {#1, ApJS, {\bf #2}, #3}
\def\aap  #1 #2 #3 {#1, A\&A, {\bf #2}, #3}
\def\mnras #1 #2 #3 {#1, MNRAS, {\bf #2}, #3}
\def\pra #1 #2 #3 {#1, Phys.~Rev.~A., {\bf #2}, #3}
\def\prb #1 #2 #3 {#1, Phys.~Rev.~B., {\bf #2}, #3}
\def\prc #1 #2 #3 {#1, Phys.~Rev.~C., {\bf #2}, #3}
\def\prd #1 #2 #3 {#1, Phys.~Rev.~D., {\bf #2}, #3}
\def\pre #1 #2 #3 {#1, Phys.~Rev.~E., {\bf #2}, #3}
\def\prl #1 #2 #3 {#1, Phys.~Rev.~Lett., {\bf #2}, #3}
\def\plb #1 #2 #3 {#1, Phys.~Lett.~B., {\bf #2}, #3}
\def\science #1 #2 #3 {#1, Science., {\bf #2}, #3}
\def\nature #1 #2 #3 {#1, Nature., {\bf #2}, #3}
\def\nphysa #1 #2 #3 {#1, Nucl.~Phys.~A., {\bf #2}, #3}
\def\nphysb #1 #2 #3 {#1, Nucl.~Phys.~B., {\bf #2}, #3}
\def\nphysbs #1 #2 #3 {#1, Nucl.~Phys.~B.~Suppl., {\bf #2}, #3}

\def\h#1{\hbox{${}^{#1}$H}}

\def\h502{\hbox{$ h^{2}_{50}$}}

\def\fun#1#2{\lower3.6pt\vbox{\baselineskip0pt\lineskip.9pt
  \ialign{$\mathsurround=0pt#1\hfil##\hfil$\crcr#2\crcr\sim\crcr}}}
\newcommand{\chic}{\chi_{+}}

\newcommand{\chif}{\chi_{0}}

\begin{document}
%
\title{Causal Structure and Gravitational Waves in Brane World Cosmology}
\author{Kiyotomo Ichiki$^{1,2}$\footnote{E-mail address: ichiki@th.nao.ac.jp}
  and Kouji Nakamura$^{3,4}$\footnote{E-mail address: kouchan@th.nao.ac.jp} 
}
\affiliation{%
  $^1$Department of Astronomy, University of Tokyo, 7-3-1 Hongo,
  Bunkyo-ku, Tokyo 113-0033, Japan}
\affiliation{%
  $^2$Division of Theoretical Astrophysics, National Astronomical
  Observatory, 2-21-1, Osawa, Mitaka, Tokyo 181-8588, Japan}
\affiliation{%
  $^3$Department of Astronomical Science, the Graduate University for
  Advanced Studies, 2-21-1, Osawa, Mitaka, Tokyo 181-8588, Japan}
\affiliation{%
  $^4$Department of Physics, Hiyoshi Campus,
  Keio University, Hiyoshi Yokohama, 223-8521, Japan
}
\date{\today}
\begin{abstract}
The causal structure of the flat brane universe of RSII type is
re-investigated to clarify the boundary conditions for stochastic
gravitational waves. 
In terms of the Gaussian normal coordinate of the brane, a singularity
of the equation for gravitational waves appears in the bulk. 
We show that this singularity corresponds to the ``seam singularity''
which is a singular subspace on the brane universe. 
Based upon the causal structure, we discuss the boundary conditions
for gravitational waves in the bulk.
Introducing a null coordinate, we propose a numerical procedure to
solve gravitational waves with appropriate boundary conditions
and show some examples of our numerical results.
This procedure could be also applied in scalar type perturbations. 
The problem in the choice of the initial condition for gravitational
waves is briefly discussed. 
\end{abstract}

\pacs{ 04.50.+h, 98.80.Cq, 98.80.-k}
\maketitle

\section{Introduction}
\label{sec:intro}


Since the proposal of a brane world model of our spacetime by Randall
and Sundrum~\cite{Randall:1999vf} (RS), the phenomenology of brane
world cosmological models has been the subject of intensive
investigations in recent years. 
In these brane world models, our universe is regarded as a four
dimensional boundary (brane) in a higher dimensional spacetime
(bulk). 
Many authors have found more realistic models which include
matter fields on the brane and realize the cosmic
expansion~\cite{Binetruy:1999hy}, and tried to constrain on these
models by the observational data.


Due to the recent developments in the technology of astronomy,
cosmological density perturbations and their observations through
large scale structure and cosmic microwave anisotropies (CMB) have
become the most stringent test to constrain on models beyond standard
cosmologies.
In order to examine the constraints on the brane cosmologies by
the CMB observations, we have to know the 5-D informations about
perturbations on the entire spacetime including the
bulk~\cite{Kodama:2000fa,Langlois:2000iu}.


In addition to the constraint by the CMB, the
stochastic gravitational waves will be a promising candidate which
provides a direct and even deeper test of such cosmologies.
One might see in principle earlier universe than the photon last
scattering epochs by gravitational waves. 
Although equations for gravitational waves are simpler than those
of density perturbations, the problem is essentially the same, i.e,
one needs to clarify the evolution of gravitational waves not only on
the brane but also in the bulk. 
Therefore, it is also instructive to clarify the minimum information
to obtain the evolution of gravitational waves before trying to
clarify that of density perturbations in brane cosmologies.


In order to give theoretical predictions of stochastic
gravitational waves, many authors have adopted the Gaussian normal
(GN) coordinate system in the neighborhood of the brane. 
In this GN coordinate system, the metric is given by the form
\begin{equation}
  ds^2 = -\frac{\psi^2(\tau,w)}{\varphi(\tau,w)}d\tau^2 
  + \varphi(\tau,w)a^{2}(\tau)d\Sigma_K^2+dw^2, 
  \label{eq:GN-coordinte-metric}
\end{equation}
and the equation of gravitational waves has the same form as that of
the five-dimensional massless scalar field:
\begin{equation}
  \Box_{5}h = 0.
  \label{eq:GN-GW-eq}
\end{equation}
The explicit forms of functions in the metric are given in the main
text (see Sec.~\ref{sec:GN-in-bulk}).
To obtain the theoretical spectrum of stochastic gravitational waves,
we just solve this equation with appropriate boundary conditions.
However, Eq.~(\ref{eq:GN-GW-eq}) in terms of the coordinate system
(\ref{eq:GN-coordinte-metric}) includes a singularity at $w=w_h$ in
the bulk, where the metric function $\varphi$ vanishes. 
The treatment of Eq.~(\ref{eq:GN-GW-eq}) near the singularity
is one of difficulties when we obtain the evolution of stochastic
gravitational waves~\cite{Hiramatsu:2003iz}.
In fact, this singularity corresponds to the ``seam singularity''
discussed by Ishihara~\cite{Ishihara:2001qe}.


The aim of this paper is to propose a numerical procedure to solve the
evolution of cosmological gravitational waves avoiding the above
difficulty in GN coordinate system. 
The procedure proposed here is based on the characteristic initial
value problem according to the causal structure of the entire
spacetime.
This idea is analogous to the analysis of gravitational waves from
a non-spherical domain wall by the one of the authors~\cite{Nakamura:2002bz}. 
We use a null coordinate instead of the proper time on the brane. 
In this procedure, the boundary conditions in the bulk are replaced by
the initial condition on a null hypersurface and the above difficulty
in the treatment of Eq.~(\ref{eq:GN-GW-eq}) near the singularity
$w=w_{h}$ is resolved if we simply specify the initial spectrum on a
null hypersurface.


The organization of this paper is as follows: 
In Sec.~\ref{sec:GN-in-bulk}, we briefly review the global structure
of a brane world universe and clarify the region covered by the GN
coordinate in terms of the closed chart of the five-dimensional
anti-de Sitter spacetime (AdS$_{5}$).
In Sec.~\ref{sec:null-causality}, we discuss the null hypersurface to
clarify the causality of the propagation of gravitational waves in the
bulk.
In Sec.~\ref{sec:eq-of-GW}, we develop the formulation to obtain the
numerical solutions to Eq.~(\ref{eq:GN-GW-eq}) and show numerical
examples of solutions which are derived by this formulation.
The final section (Sec.~\ref{sec:summary-discussion}) is devoted to
summary and discussions.


Throughout this paper, we consider the model without ``dark
radiation'' following discussions in Ref.\cite{Ichiki:2002eh} and
we only consider the flat Friedmann-Robertson-Walker (FRW) brane
universe which is supported by recent precise measurements of the
CMB~\cite{Spergel:2003cb}.


\section{GN coordinate system in ${\bf AdS_{5}}$ bulk}
\label{sec:GN-in-bulk}


In this section, we first relate  GN coordinate 
system to the flat chart in AdS$_{5}$, and then consider the 
correspondence of GN coordinate system and the closed chart in AdS$_{5}$ 
which covers the entire AdS$_{5}$.
Through clarifying these relations, we can easily see that the region 
which is covered by GN coordinate system in the entire AdS$_{5}$. 
Secondly, we find where the singularity of the equation
(\ref{eq:GN-GW-eq}) is in the bulk.


Now, we consider a brane universe embedded in the AdS$_{5}$
with a negative cosmological constant $\Lambda_{5}=-4/l^{2}$. 
In terms of the static charts of AdS$_{5}$, the metric on AdS$_{5}$ is
given by  
\begin{equation}
 ds^2=-f_K(r_K)dt_K^2+f_K(r_K)^{-1}dr_K^2+r_K^2d\Sigma_K^2 ,
  \label{rtmetric}
\end{equation}
where $K$ takes the values $-1, 0,$ and $+1$, corresponding to the 
negative, zero, and positive constant curvature of a maximally
symmetric three dimensional space, respectively (for example, see
Ref.\cite{Ishihara:2001qe}).
The function $f_{K}(r_K)$ in this metric is defined by 
\begin{equation}
 f_K(r_K):=K+\frac{r_K^2}{l^2}.
\end{equation}
When $K=+1$ and $0$, the metric $d\Sigma_K^2$ is given by 
\begin{eqnarray}
  d\Sigma_K^2 = \left\{
    \begin{array}{ll}\displaystyle
         d\chic^2 + \sin^2{\chic}\{d\theta^2 + \sin^2{\theta}d\phi^2\}
& (K=+1),  \cr
         d\chif^2 + \chif^2\{d\theta^2 + \sin^2{\theta}d\phi^2\}
& (K=0), 
    \end{array}
  \right.
\nonumber
\end{eqnarray}
respectively.
Though we may consider the case $K=-1$ which corresponds to the open
FRW model, we do not treat this case in this paper.
In these static charts, a trajectory of three brane is given by
$r_{K}=r_{K}(t_{K}):=a(\tau)$ and $t_{K}=t_{K}(\tau)$, where $\tau$ is
the proper time of the world volume of the brane and $a(\tau)$ is a
cosmological scale factor on the brane.
The equation of the brane motion is given by the generalized
Friedmann equation~\cite{Binetruy:1999hy}
\begin{equation}
  H^2 = \left(\frac{\dot{a}}{a}\right)^2
  =\frac{8 \pi G_{\rm N}}{3}\rho
  -\frac{K}{a^2}+\frac{\Lambda_{4}}{3}
  +\frac{\kappa_{5}^4}{36}\rho^2, 
  \label{Friedmann}
\end{equation}
where the dot denotes the derivative with respect to $\tau$, $\rho$ is
the energy density on the brane, 
$\Lambda_{4}=\kappa_{5}^4 \lambda^2 /12 + 3 \Lambda_{5}/4$ is the 
cosmological constant induced on the brane, $G_{\rm N}=\kappa_{5}^4
\lambda / 48 \pi$ and $\kappa_{5}$ are the four-dimensional and
five-dimensional gravitational constants, respectively. Here and
hereafter we assume $Z_2$ symmetry across the brane.


Cosmological solutions of the RS type brane world in
terms of GN coordinate system were found by several authors
~\cite{Binetruy:1999hy}, which is given by
Eq.~(\ref{eq:GN-coordinte-metric}).
In this metric, functions $\psi$ and $\varphi$ are given by 
\begin{eqnarray}
  \varphi(\tau,w)=
  \frac{1}{4}\left[(1-A)e^\frac{w}{l}+(A+1)e^{-\frac{w}{l}}\right]^2 
  \label{eq:varphi-explicit}
\end{eqnarray}
and
\begin{equation}
  \psi(\tau,w)=\varphi(\tau,w)+\frac{1}{2H}\left(\frac{\partial \varphi}{\partial \tau}\right)_{w,x^i} ,
\end{equation}
where $A:=\sqrt{1+l^2H^2}$.
We have chosen the coordinate $w$ so that $w=0$ on the brane.
As noted in Sec.~\ref{sec:intro}, this coordinate
system has a coordinate singularity at $w=w_h$, where $w_h$ is
determined by the equation $\varphi(\tau,w_h)=0$ for each $\tau$.
This equation yields
\begin{eqnarray}
  \exp\left(2\frac{w_h}{l}\right)&=& \frac{A+1}{A-1} .
\end{eqnarray}
This is nothing but the singularity in Eq.~(\ref{eq:GN-GW-eq}).


To give the explicit coordinate transformations between the metrics
(\ref{eq:GN-coordinte-metric}) and (\ref{rtmetric}) with $K=0$,
we show the explicit forms of the functions $(t_{0},r_{0})$ in terms
of $(\tau,w)$.
The function $r_{0}$ in Eq.~(\ref{rtmetric}) corresponds to the
volume element on the $\tau=const.$ hypersurface in the brane.
Then, as shown in Ref.\cite{Mukohyama:1999wi}, this function is given
by
\begin{eqnarray}
  r_0^2&=&\varphi(\tau,w)a^2(\tau).
  \label{r_0oftauw}
\end{eqnarray}
The cosmological scale factor $a(\tau)$ is given by
$a(\tau):=r_{0}(\tau,w=0)$.  
%
%
%
%
On the other hand, the explicit expression of $t_0$ as a function
of ($\tau$,$w$) is given by 
\begin{eqnarray}
  t_0-t_b
  =\frac{l^{2}}{a}\sqrt{\frac{A+1}{A-1}}\left[1
    -\frac{2}{(1-A)e^{2\frac{w}{l}}+1+A}\right],
  \label{t0astauw}
\end{eqnarray}
where $t_b(\tau)$ is chosen so that $t_{0}=t_{b}(\tau)$ on the brane
($w=0$) for any $\tau$.
The derivation of this expression (\ref{t0astauw}) is shown in
Appendix \ref{sec:t_0-derivation}.
%
%
%
%
Clearly, the singularity in the equation (\ref{eq:GN-GW-eq}) is just 
on the region $r_0=0$ in the flat chart.
Actually, in the vicinity of the singularity ($w \sim w_h$), $r_{0}$
and $t_{0}$ behave as 
\begin{eqnarray}
 r_0 &=&
 \frac{a(\tau)}{2}\sqrt{\frac{A-1}{A+1}}\left(\xi+\frac{1}{2(A+1)}\xi^{2}\right)
 + O(\xi^{3}), \\
 t_0 &=& t_{b} 
 + \frac{l^2}{a(\tau)}\sqrt{\frac{A+1}{A-1}} \left(
   - \frac{2}{\xi} + 1
   \right),
\end{eqnarray}
where $\xi:=(1-A)e^{2\frac{w}{l}}+1+A$.
To find where is the singularity in Eq.~(\ref{eq:GN-GW-eq})
in the bulk, we first see the region of $r_0=0$ in the entire
AdS$_{5}$ using the closed chart.
On this closed chart, we can easily specify the point of the
singularity in Eq.~(\ref{eq:GN-GW-eq}) in AdS$_{5}$ bulk by tracing
the spacelike geodesic which normal to the brane for each $\tau$.


Now, we consider the relation between the flat chart and the closed
chart of AdS$_{5}$.
Though this is already given by Ishihara\cite{Ishihara:2001qe}, we
repeat his arguments to find the event $\varphi(\tau,w_h)=0$ for each
$\tau$ in the closed chart.
The AdS$_{5}$ is identified with the universal covering space of a
hyperboloid, 
\begin{equation}
 -Y_0^2-Y_1^2 +\sum_{i=2}^5 Y_i^2=-l^2~~,
\end{equation}
in six-dimensional flat spacetime with the metric,
\begin{equation}
 ds^2=-dY_0^2-dY_1^2+\sum_{i=2}^5 dY_i^2~~.
\end{equation}
The flat chart and the closed chart (Eq.~(\ref{rtmetric})) cover this 
hyperboloid as follows: For the $K=+1$ case (the closed chart),
\begin{equation}
 \left(
  \begin{array}{c}
   Y_0 \\
   Y_1 \\
   Y_2 \\
   Y_3 \\
   Y_4 \\
   Y_5
  \end{array}
\right)=\left(
\begin{array}{c}
   \sqrt{l^2+r_+^2}\sin(t_+/l) \\
   \sqrt{l^2+r_+^2}\cos(t_+/l) \\
   r_+\cos\chi_+ \\
   r_+\sin\chi_+\cos\theta \\
   r_+\sin\chi_+\sin\theta\cos\phi \\   
   r_+\sin\chi_+\sin\theta\sin\phi
\end{array}
\right)~~;
\label{closedchart}
\end{equation}
For the $K=0$ case (the flat chart),
\begin{equation}
 \left(
  \begin{array}{c}
   Y_2+Y_0 \\
   Y_2-Y_0 \\
   Y_1 \\
   Y_3 \\
   Y_4 \\
   Y_5
  \end{array}
\right)=\left(
\begin{array}{c}
   r_0 \\
   \left(\frac{t_0^2}{l^2}-\chi_0^2\right)r_0-\frac{l^2}{r_0}\\
   -\frac{t_0}{l}r_0\\
   r_0\chi_0\cos\theta \\
   r_0\chi_0\sin\theta\cos\phi \\
   r_0\chi_0\sin\theta\sin\phi \\
\end{array}
\right)~~.
\label{flatchart}
\end{equation}
We should note that the flat chart does not cover the entire AdS$_{5}$  
since $Y_2+Y_0 \ge 0$ ($r_{0}\ge 0$). 
Comparing Eqs.~(\ref{closedchart}) with Eqs.~(\ref{flatchart}), we
found the coordinate transformations from the flat chart to the closed
ones: 
\begin{eqnarray}
  && r_+^2=\frac{1}{4}\left(r_0-\frac{l^2}{r_0}
                +\left(\frac{t_0^2}{l^2}-\chi_0^2\right)r_0\right)^2+r_0^2\chi_0^2 ~~,
                     \nonumber \\
  &&\cos \chi_+=\frac{r_0}{2r_+}\left[1+\frac{t_0^2}{l^2}-\chi_0^2-\frac{l^2}{r_0^2}\right]~~,
 \nonumber \\
 && \sin \chi_+=\frac{r_0\chi_0}{r_+}~~,\label{flattoclosed} \\
 && \tan\left(\frac{t_+}{l}-\pi\Theta(t_{0})\right)=-\frac{l}{2t_0}\left(1+\frac{l^2}{r_0^2}
               -\frac{t_0^2}{l^2}+\chi_{0}^2\right)~~,\nonumber
\end{eqnarray}
where $\Theta(t_{0})$ is the step function, i.e., 
$\Theta(t_{0})=0$ for $t_{0}\leq 0$ and 
$\Theta(t_{0})=1$ for $t_{0}> 0$.


\begin{figure}[htb]
\rotatebox{-90}{\includegraphics[width=0.22\textwidth]{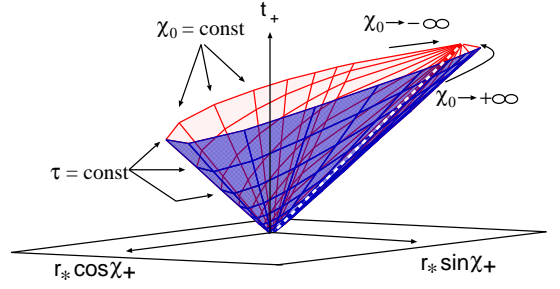}}
\caption{Embedded flat FRW brane. Hypersurfaces at which $\tau$ or
 $\chi_0$ are constant are also shown in the figure. }
\label{earlyembedded.v2.eps}
\end{figure}


In addition to these coordinate systems, it is convenient to introduce
the temporal-radial subspace which makes the metric conformally flat  
\begin{eqnarray}
 ds^2&=&f_+(r_+)(-dt_+^2+dr_\ast^2)+r_+^2 d\Sigma_+^2, \\
 (r_\ast/l)&=&\arctan(r_+/l) 
\end{eqnarray}
for the purpose of the investigation of the causal structure. 
Note that in these coordinates, radial null rays are
represented by straight lines at $\pm 45$ degree.
In the coordinate system $(t_+,r_\ast \cos\chi_+,r_\ast \sin \chi_+)$,
the flat FRW brane is embedded as shown in Fig.~\ref{earlyembedded.v2.eps}. 
The $\tau=const$ and $\chi_{0}=const$ hypersurfaces on the flat FRW
brane is also shown in the same figure, respectively.

\begin{figure}[ht]
\rotatebox{0}{\includegraphics[width=0.30\textwidth]{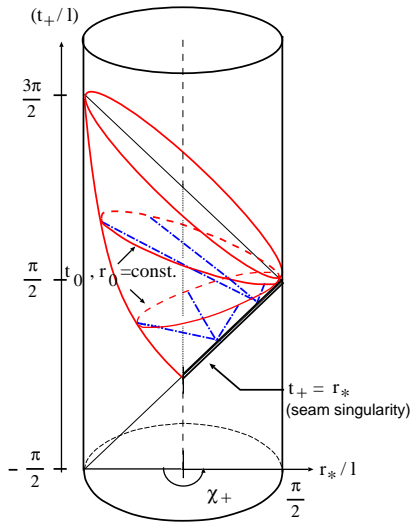}}
\caption{Geodesics normal to the brane for each $\tau$ (dot-dashed
  lines) in the entire AdS$_{5}$ spacetime. The geodesics converge to
 the seam singularity $t_+=r_*$ at which $\varphi(\tau,w)=0$.}
\label{causal.eps}
\end{figure}

Through the relations (\ref{r_0oftauw}), (\ref{t0astauw}), and
(\ref{flattoclosed}), we can see how $\tau=const.$ hypersurfaces 
in the coordinate system (\ref{eq:GN-coordinte-metric}) foliate the
entire AdS$_{5}$ bulk. 
To do this, we have to consider the spacelike geodesics 
normal to the brane specifying their starting point on the brane. 
Note that on each $\tau=const.$ hypersurface in the brane ($w=0$),
both functions $t_{0}(\tau)$ and $r_{0}(\tau)$ are also constant,
while the other coordinates on the brane are arbitrary.
Therefore, we may choose a point $(t_{0},r_{0})$ on the flat chart as
a starting point of these spacelike geodesics normal to the brane.
Fixing the proper time $\tau$ on the brane,
the coordinate functions $r_{+}$, $t_{+}$, and $\chi_{+}$ behave as 
\begin{eqnarray}
 r_+ &=& \frac{l^2}{a(\tau)}\frac{A}{\sqrt{A^2-1}}+t_b, \nonumber \\
 \frac{t_+}{l} &=& \arctan\left[\frac{r_+}{l}\right] = \frac{r_{*}}{l}, \\
 \cos \chi_+ &=& -1 \nonumber
\end{eqnarray}
in the limit $w \to w_h$ ($\varphi(\tau,w)\to 0$).
This shows that the coordinate singularity at which
$\varphi(\tau,w_{h})=0$ is just on the ``seam
singularity'' $t_+=r_{*}$, $\cos\chi_+=-1$ (see Fig.~\ref{causal.eps}).


Note that the seam singularity can be replaced by
the other regular portion of a spacetime according to the creation
scenario of the brane universe. 
In this sense, we do not have to be afraid of this singularity, or
equivalently, the singularity in Eq.~(\ref{eq:GN-GW-eq}), seriously.
Moreover, when we consider the characteristic initial value problem, 
the initial conditions are chosen on a null hypersurface and the seam
singularity is contained in its causal past.
Therefore, the difficulties in the treatment of the singularity in
Eq.~(\ref{eq:GN-GW-eq}) are simply reduced to the choice of initial
conditions for gravitational waves as seen in below.


\section{Null Hypersurface and Causality in the Bulk}
\label{sec:null-causality}


To clarify the causal
structure of the brane universe and the domain of dependence of 
gravitational waves, let us now consider the future directed radial null
geodesics from the brane into the bulk. 
These null geodesics determine the boundary associated with the causal
region of the events on the brane.


In terms of the flat chart, the equations of these null geodesics 
are given by
\begin{eqnarray}
 \frac{dt_0}{d\lambda}=\frac{l^2}{r_0^2}E, \quad
 \frac{dr_0}{d\lambda}= - E, \quad
 \frac{d\chi_0}{d\lambda}=0,
 \label{eq:radial-null-geodesics-eq}
\end{eqnarray}
where $E (>0)$ is an integration constant and $\lambda$ is the affine
parameter along the geodesics.
Because we concentrate on the radial null geodesics from the brane, $\chi_{0}$
is constant along the geodesics and the remaining equations in
Eqs.~(\ref{eq:radial-null-geodesics-eq}) yield
\begin{eqnarray}
 t_0=\frac{l^{2}E\lambda}{r_b(-E\lambda+r_b)}+t_b, \quad
 r_0=-E\lambda+r_b \label{nullinflat}.
\end{eqnarray}
In Eqs.~(\ref{nullinflat}), $t_b$ and $r_b$ are constants which
correspond to the initial condition for the null geodesics. 
We regard that the point $(t_b,r_b)$ is on the brane in the flat chart 
and this point is the starting point of the geodesics.
For this reason, we regard $t_b$ and $r_b$ as functions of the
proper time $\tau$ on the brane and these functions are constrained by
the generalized Friedmann equation Eq.~(\ref{Friedmann}).


Here, we note that the flat chart of AdS$_{5}$ has the future Cauchy
horizon in the entire AdS$_{5}$. 
This Cauchy horizon appears as the ``infinity'' 
$(t_0,r_0)\to(+\infty,+0)$ in the flat chart. 
The above radial null geodesics from the brane approaches to this
Cauchy horizon as the affine parameter $\lambda$ increases.
We choose the affine parameter $\lambda$ so that $\lambda=0$
corresponds to the point on the brane and $\lambda=\frac{E}{r_b}>0$
corresponds to that on the Cauchy horizon.


On the other hand, in the closed chart, these null geodesics are given
by 
\begin{eqnarray}
 && r_+^2 =
 \frac{1}{4}\left[\left(1+\frac{\zeta^2}{l^2}-\chi_0^2\right)r_0
 +2\zeta\right]^2+r_0^2\chi_0^2, \nonumber \\
 && \tan\left(\frac{t_+}{l}-\pi\right)=\frac{l}{l^2+\zeta
 r_0}\left\{\zeta-\frac{1}{2}\left(1-\frac{\zeta^2}{l^2}+\chi_0^2\right)r_0
      \right\}, \nonumber \\
 && \cos \chi_+ =
 \frac{1}{2r_+}\left[\left(1+\frac{\zeta^2}{l^2}-\chi_0^2\right)r_0+2\zeta\right],
 \label{eq:null-geo-closed}
\end{eqnarray}
where $\zeta := -\frac{l^2}{r_b}+t_b$.
In Eqs.~(\ref{eq:null-geo-closed}), $t_0$ and $r_0$ are parameterized
by $\lambda$ through Eqs.~(\ref{nullinflat}).
In the limit where the null geodesics from the brane approach to
the Cauchy horizon $(t_0,r_0) \to (+\infty,+0)$,
Eqs.~(\ref{eq:null-geo-closed}) behave
\begin{eqnarray}
 r_+= |\zeta|, \;\;
 \frac{t_+}{l}=\arctan\left[\frac{\zeta}{l}\right]+\pi, \;\;
 \cos \chi_+ = {\rm sgn}(\zeta). 
 \label{futurenull}
\end{eqnarray}
These do represent a point on the future Cauchy horizon of the flat
chart.


\begin{figure}[htb]
\rotatebox{0}{\includegraphics[width=0.27\textwidth]{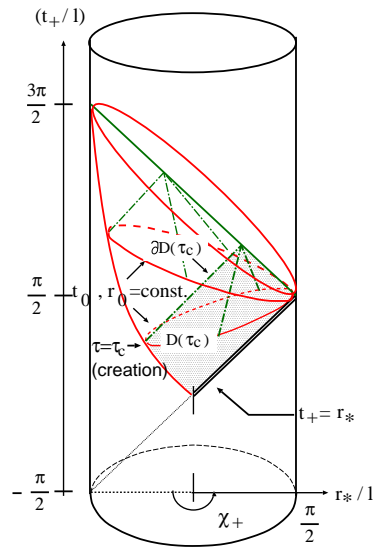}}
\caption{Brane trajectory (lines) and future null geodesics from the
  brane (dot-dashed lines). Singularity where $\varphi(\tau,w_h)=0$ is
  shown as a thick line, which is entirely contained in the shaded
 region ${\cal D}(\tau_c)$.
 The evolution of gravitational waves is determined by boundary
 conditions on the brane and an initial spectrum on $\partial {\cal D}(\tau_c)$.
The initial spectrum is determined by the creation scenario of the brane
 universe.}
\label{causal2.eps}
\end{figure}


Actually, from the closed chart metric (\ref{rtmetric}) with $K=+1$,
we can easily see that the event represented by
Eqs.~(\ref{futurenull}) is pointlike when we fix the proper time
$\tau$ on the brane. 
We also note the fact that Eqs.~(\ref{futurenull}) do not depend on
$\chi_{0}$.
This implies that all null geodesics with different $\chi_{0}$ from
the same $\tau=const.$ hypersurface on the brane reach to the same
point represented by Eqs.~(\ref{futurenull}).
These null geodesics generate a null hypersurface.
Hence we have one-to-one correspondence between a $\tau=const.$
hypersurface on the brane and a null hypersurface in the bulk.
We denote this null hypersurface associated with $\tau=const.$ hypersurface on the brane by $\partial{\cal D}(\tau)$.
We also denote the causal
past of $\partial {\cal D}(\tau)$ by ${\cal D}(\tau)$.
We note that ${\cal D}(\tau')\subset{\cal D}(\tau)$ for any
$\tau'<\tau$. 
In particular, the bulk region covered by the flat chart is foliated
by the set of null hypersurfaces \{$\partial {\cal D}(\tau)$ $|$ $\tau$ is
the proper time on the brane\}. Since $\partial {\cal D}(\tau')\subset
{\cal D}(\tau)$ for $\tau'<\tau$, the seam singularity, which is also
represented by $\partial {\cal D}(\tau \to +0)$, is entirely contained
in ${\cal D(\tau)}$ for any $\tau$.
These situations are schematically depicted in
Fig.~\ref{causal2.eps}.


Inspecting this causal structure, we can formulate the characteristic
initial value problem of gravitational waves.
Suppose that the flat FRW brane universe is created at the
instance $\tau=\tau_{c}$.
As depicted in Fig.~\ref{causal2.eps}, the initial state of
gravitational waves in ${\cal D}(\tau_{c})$ affects to their evolution
only through the null hypersurface $\partial{\cal D}(\tau_{c})$. 
Once we specify the state of gravitational waves in the region 
${\cal D}(\tau_{c})$ according to the creation scenario of the flat
FRW brane, we can specify the spectrum of gravitational waves on
$\partial{\cal D}(\tau_{c})$. 
Hence, to determine the evolution of the spectrum of gravitational waves
after $\tau=\tau_{c}$, we only have to specify the
spectrum of gravitational waves on $\partial{\cal D}(\tau_{c})$ as an
initial condition and trace the 
evolution of them through Eq.~(\ref{eq:GN-GW-eq}) as seen in the next
section.


\section{Evolution of Gravitational Waves}
\label{sec:eq-of-GW}


At this point, we have a clear strategy to tackle the problem of the
evolution of gravitational waves in the brane universe.
The singularities $\varphi(\tau,w_{h})=0$ in Eq.~(\ref{eq:GN-GW-eq})
are entirely contained in ${\cal D}(\tau)$, and the evolution of gravitational
waves are determined by an initial spectrum on its boundary
$\partial {\cal D}(\tau)$ and boundary conditions on the brane.
Then, we do not have to care the singularity $\varphi(\tau,w_{h})=0$
any more and there is no need to introduce any artificial boundaries in
the bulk to impose boundary conditions at bulk infinity.
In the following subsections, we first develope the formulation to
obtain the numerical solutions to Eq.~(\ref{eq:GN-GW-eq}) in the
context of brane world cosmology (in Sec.~\ref{sec:eq-of-GW-sub-1}), 
and then, we show numerical examples which are obtained by applying the
formulation developed here (Sec.~\ref{sec:eq-of-GW-sub-2}).


\subsection{Numerical Formulation}
\label{sec:eq-of-GW-sub-1}

\begin{figure}[h]
  \begin{center}
    \rotatebox{-90}{\includegraphics[width=0.35\textwidth]{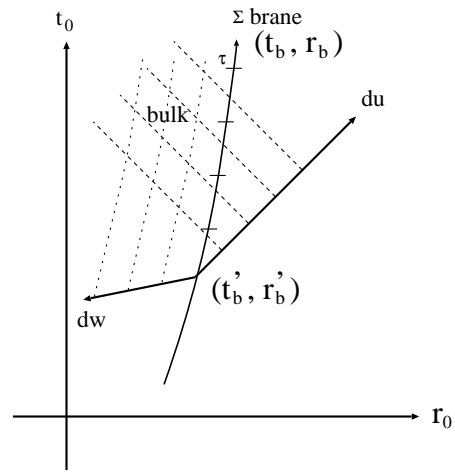}}
    \caption{Null coordinate and Gaussian normal coordinate in the flat chart.}
    \label{nullcoordinate.eps}
  \end{center}
\end{figure}

In order to develop the characteristic initial value problem
associated with the null hypersurface $\partial{\cal D}(\tau)$, we
introduce a null coordinate and rewrite down the equation of
gravitational waves in a single null coordinate system (see
Fig.~\ref{nullcoordinate.eps}).


Now, we introduce the function $u$ by 
\begin{eqnarray}
  u=t_0-\frac{l^2}{r_0}-t_{b}^\prime+\frac{l^2}{r_{b}^\prime}~~,
\label{u_of_t0_r0}
\end{eqnarray}
where ($t_{b}^\prime$,$r_{b}^\prime$) determine the zero point of this
function $u$.
The function $u$ is constant on the null hypersurface $\partial{\cal D}(\tau)$ for
each $\tau$. 
We can easily confirm that the one-form $(du)_{a}$ is null through the
metric (\ref{rtmetric}) with $K=0$.


\vspace*{2pc}


Using the relations listed in the appendix (Eq.(\ref{A6})-(\ref{A9}),
see also Ref.~\cite{Mukohyama:1999wi}), 
the derivative $du$ is given by 
\begin{eqnarray}
  du &=& \left(\frac{\partial u}{\partial \tau}\right)_w d\tau +
  \left(\frac{\partial u}{\partial w}\right)_\tau dw \nonumber\\ 
  &=& \frac{1}{\dot a}\left(\frac{\partial r_0}{\partial \tau}\right)_w F^{-1}(\tau,w)d\tau-F^{-1}(\tau,w)dw~~,
\label{dudtaudw}
\end{eqnarray}
where
\begin{equation}
 F(\tau,w)=\frac{f[r_0(\tau,w)]}{\sqrt{f[r_0(\tau,w)]+{\dot a}^2}+\dot a}~~.
\end{equation}
In terms of the new coordinate system $u$ and ${\overline w}=w$, the
five-dimensional metric (\ref{eq:GN-coordinte-metric}) on AdS$_{5}$ is
written as
\begin{eqnarray}
  ds^2 = - F(\tau,{\overline w})du^2
  - 2F(\tau,{\overline w})dud{\overline w}
  + r_0^2(\tau,{\overline w})d\Sigma_0^2, 
  \label{nullmetric}
\end{eqnarray}
As mentioned in Sec.~\ref{sec:intro}, the equation
for gravitational waves in the bulk is simply given by that for the
five-dimensional massless scalar field Eq.~(\ref{eq:GN-GW-eq}).
In terms of the coordinate system $(u,{\overline w})$
Eq.~(\ref{eq:GN-GW-eq}) is given by
\begin{widetext}
\begin{equation}
  \left[
    \partial_{\overline w}^2-\frac{2}{F}\partial_{\overline w} \partial_u
    -\frac{3}{Fr_{0}}\left(
      \left(\frac{\partial r_{0}}{\partial u}\right)_{{\overline w}}
      \partial_{\overline w}
      +\left(\frac{\partial r_{0}}{\partial \overline w}\right)_{u}
        \partial_u \right)
    +\left(\frac{1}{F}
      \left(\frac{\partial F}{\partial {\overline w}}\right)_{u}
      +\frac{3}{r_{0}}
      \left(\frac{\partial r_{0}}{\partial {\overline w}}\right)_{u}
      \right)
    \partial_{\overline w}
    -\frac{k^2}{r^2_{0}}\right] h(u,\overline w;k)=0~~,
  \label{gweq}
\end{equation}
\end{widetext}
where $-k^{2}$ is the eigen value of the Laplacian of $d\Sigma_0$.


Eq.~(\ref{gweq}) can be formally integrated so that  
\begin{eqnarray}
  \partial_uh(u,\overline w)&=&r^{-3/2}_{0}(\tau,\overline w) \nonumber \\
  &\times&\left(\int^{{\overline w}}S(w^\prime) 
    r^{3/2}_{0}dw^\prime+C(u)\right) ,
\label{intsolution}
\end{eqnarray}
where
\begin{eqnarray}
  S({\overline w})&=&\frac{F}{2}\left(\partial_{\overline w}^2h
    - \frac{k^2}{r^2_{0}}h\right) \nonumber \\
  &+&\left(\frac{1}{2}\frac{\partial F}{\partial \overline w}
    +\frac{3F}{2}\frac{\partial \ln r_{0}}{\partial \overline w}
    -\frac{3}{2}\frac{\partial \ln r_{0}}{\partial u}\right) 
  \partial_{\overline w} h~~.
\label{S(w)}
\end{eqnarray}
The quantities $\partial r_{0}/\partial u$, 
$\partial r_{0}/\partial{\overline w}$ and 
$\partial F/\partial{\overline w}$ 
which appear in the definition (\ref{S(w)}) are
given by known functions as
\begin{eqnarray}
  \left(\frac{\partial r_{0}}{\partial u}\right)_{{\overline w}} 
  &=& \dot a F~~,
  \label{eq:pr0overpu}
  \\
  \left(\frac{\partial r_{0}}{\partial \overline w}\right)_{u} 
  &=& \dot a +\left(\frac{\partial r_{0}}{\partial w}\right)_\tau ~~,
  \label{eq:pr0overwbar}
  \\ 
  \left(\frac{\partial F}{\partial \overline w}\right)_{u} 
  &=& \dot a\left(\frac{\partial r_{0}}{\partial
      \tau}\right)_w^{-1}\left(\frac{\partial F}{\partial \tau}\right)_w
  + \left(\frac{\partial F}{\partial w}\right)_\tau~~.
  \label{eq:pFoverwbar}
\end{eqnarray}
These equations (\ref{intsolution}) and (\ref{S(w)}) are the main
result of this paper. 
In these equations, initial data of gravitational waves are set by
choosing the function of $h(u,{\overline w})$ on $\partial {\cal
D}(\tau)$ where the null coordinate $u$ is constant on $\partial
{\cal D}(\tau)$.
The function $C(u)$ in Eq.~(\ref{intsolution}) is determined by the
boundary condition at the brane ($\overline w =0$) by 
\begin{equation}
  \partial_u h(u,0) = r^{-3/2}_{0}(\tau,0) C(u) .
\end{equation}
When the anisotropic stress on the brane is not induced due to matter
fields on the brane, the boundary condition for gravitational waves
$h$ at the brane is the Neumann type~\cite{Kodama:2000fa}. 
This is accomplished by imposing 
\begin{eqnarray}
 \partial_uh(u,0)&=&-\left(\frac{\partial u}{\partial w}\right)^{-1}_\tau \partial_{\overline w}h \nonumber \\
 &=&F(\tau,0) \partial_{\overline w}h~~,
\label{boundarycondition}
\end{eqnarray}
at the brane.
Eq.~(\ref{intsolution}) with the boundary condition
(\ref{boundarycondition}) are easier to solve than
Eq.~(\ref{eq:GN-GW-eq}) and enough to predict the cosmological
evolution of gravitational waves.
Note that our method have an important advantage that application to
high energy epochs is easy and straightforward. 
Actually, this advantage is seen in the numerical examples 
shown in the following subsection (see Figs.~\ref{fig5.eps} and
\ref{fig6.eps}).


\subsection{Numerical Scheme and Examples}
\label{sec:eq-of-GW-sub-2}


Here, we show some numerical solutions to Eq.~(\ref{eq:GN-GW-eq}) in
the context of the brane world cosmology.
We first comment on some details of the numerical scheme to obtain
solutions to Eq.~(\ref{gweq}) with the boundary condition
Eq.~(\ref{boundarycondition}).


To solve Eq.~(\ref{gweq}), we evaluate Eq.~(\ref{intsolution}) on each
null hypersurface $\partial{\cal D}(\tau(u,\overline w=0))$, numerically.
Once $h(u,\overline w)$ on $\partial{\cal D}(\tau(u,\overline w=0))$ 
is given as a function of $\overline w$, we can evaluate
$h(u+du,\overline w)$ on $\partial{\cal D}(\tau(u+du,\overline w=0))$ by 
\begin{equation}
  h(u+du,\overline w) = h(u,\overline w) + du \partial_u h(u,\overline w).
\end{equation}
Through this evaluation on each 
$\partial{\cal D}(\tau(u,\overline w=0))$, we obtain the gravitational
waves $h(u,\overline w)$ in the causal future of the initial surface
$\partial{\cal D}(\tau_{ini})$ with an appropriate initial condition.
By evaluating $h(u,0)$ on the brane, we see the behavior of
gravitational waves on our brane universe.


To obtain $\partial_u h(u,\overline w)$ by Eq.~(\ref{intsolution}), we
have to evaluate Eqs.~(\ref{eq:pr0overpu})-(\ref{eq:pFoverwbar})
on each $\partial{\cal D}(\tau(u,\overline w=0))$ and
Eq.~(\ref{boundarycondition}) at the brane.
To accomplish this, we evaluate $\dot{a}(\tau)=a(\tau)H(a(\tau))$,
$F(\tau,\overline w)$, and $r_{0}(\tau,\overline w)$ on each
$\partial {\cal D}(\tau(u,\overline w=0))$.
Since these functions depend on $u$ only through $\tau$, 
we evaluate the change of the function $\tau(u,\overline w)$ on
each $\partial {\cal D}(\tau(u,\overline w=0))$. 
Further, it is also convenient to use the scale factor $a(\tau)$ at
the brane as an time coordinate when we clarify the behavior of
gravitational waves during the cosmic expansion.


We evaluate the change of the function $a(\tau(u,\overline w))$
on $\partial{\cal D}(\tau(u,\overline w=0))$ from
Eq.~(\ref{dudtaudw}) by setting $du=0$, which yields
\begin{equation}
 \left(\frac{d\ln a}{dw}\right)_{u} =
  a H^2 \left(\frac{\partial r_0}{\partial \tau}\right)_w^{-1} ~~,
\label{dlna/dw}
\end{equation}
where $(\partial r_0/\partial \tau)_w$ is directly given by
Eq.~(\ref{r_0oftauw}). 
By integrating Eq.(\ref{dlna/dw}) with the boundary condition
$a(\tau)=r_{0}(\tau,w=0)$ at the brane, we obtain the function
$a(\tau(u,\overline w))$ along $\partial{\cal D}(\tau(u,\overline w=0))$.
Since the relation between $a$ and $\tau$ is given by the
explicit integration of the generalized Friedmann equation
Eq.~(\ref{Friedmann}), we can evaluate $F(\tau,w)$ and
$r_0(\tau,w)$ on each $\partial{\cal D}(\tau(u,\overline w=0))$ by
using Eqs.~(\ref{Friedmann}) and (\ref{r_0oftauw}). 
The relation between the time step $du$ and $d\tau$ on the brane is given by 
\begin{equation}
 d\tau = \dot a \left(\frac{\partial r_0}{\partial \tau}\right)^{-1}_w
 F(\tau,w=0) du~,
\end{equation}
which is led from Eq.~(\ref{dudtaudw}) by choosing $dw=0$ and 
$w=0$.
After proceeding the time step, we repeat the above evaluation
of $a(\tau)$ and $\tau(u,\overline w)$ on 
$\partial{\cal D}(\tau(u+du,\overline w=0)$. 
Thus, we can evaluate $F(\tau,\overline w)$,
$r_{0}(\tau,\overline w)$, and $\dot{a}$ on each null
hypersurface $\partial{\cal D}(\tau(u,\overline w=0))$ from
Eqs.~(\ref{Friedmann}), 
(\ref{r_0oftauw}), and (\ref{dudtaudw}), and hence,
 Eqs.~(\ref{eq:pr0overpu})-(\ref{eq:pFoverwbar}).
Through these evaluations, we obtain $\partial_u h(u,\overline w)$ by
Eq.~(\ref{intsolution}) and gravitational wave $h(u,\overline w)$ in
the causal future of the initial surface $\partial{\cal D}(\tau_{ini})$.


When we carry out this numerical scheme, we have to specify the zero
point of the function $u$.
Suppose that we start our numerical calculation from the high
energy epoch ($l^2H^2\gg1$, or the $\rho^{2}$ dominated era).
In this epoch, the generalized Friedmann equation and
Eq.~(\ref{dtau_dt0}) lead
\begin{eqnarray}
  H^2 &\sim& \frac{1}{16} \tau^{-2}~,\\
  \frac{a}{a_0} &=& 
  \left(\frac{2}{3}\kappa_5^{2}\rho_{\gamma 0}\tau\right)^{1/4} 
  := \alpha \tau^{1/4}~, \\
  t_b(\tau)-t_{b,\rm ini}' &\sim&
  - \frac{l^2}{\alpha a_0}\left(\tau^{-1/4}-\tau_{\rm ini}^{-1/4}\right), 
\end{eqnarray}
where $\rho_{\gamma 0}$ and $a_0$ are the energy density of
radiations and the cosmic scale factor on the brane at the
present universe.
$t_{b,ini}^\prime$ is the trivial initial value of $t_b$ and we
choose so that $t_{b,\rm ini}^\prime=0$. 
Since the zero points of $u$ and $t_{b}$ are arbitrary, we choose so
that $u=t_b^\prime=0$ and $r_b^\prime = a(\tau_{\rm ini})$ in
Eq.(\ref{u_of_t0_r0}) at the starting point of our numerical
calculation $\tau=\tau_{\rm ini}=1/4H(a_{\rm ini})$.

\begin{figure}[ht]
\rotatebox{0}{\includegraphics[width=0.40\textwidth]{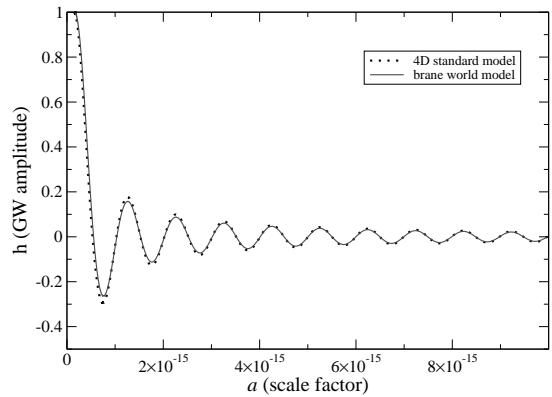}}
\caption{Evolution of gravitational waves on the brane in brane world
 cosmology (solid line) which cross the horizon in low energy epochs
 ($l^2H^2\ll1$). Also shown is standard 4D evolution of gravitational
 waves (dotted line).}
\label{fig5.eps}
\end{figure}

\begin{figure}[ht]
\rotatebox{0}{\includegraphics[width=0.40\textwidth]{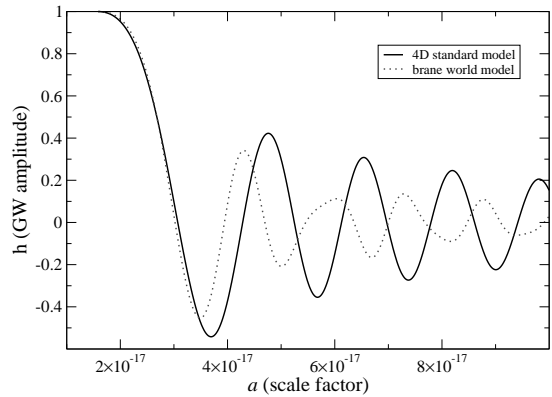}}
\caption{The same as Fig.~\ref{fig5.eps}, but for gravitational waves
  which cross the horizon in high energy epochs ($l^2H^2\gg1$).}
\label{fig6.eps}
\end{figure}

Following to these numerical scheme, we obtain numerical
examples depicted in Figs.~\ref{fig5.eps} and \ref{fig6.eps}. 
In these examples, we simply choose $h = const.$ as initial conditions
of $h$ on the initial null surface $\partial {\cal D}(\tau_{\rm ini})$.
We take this initial condition only as an example to demonstrate the
calculation and investigate how the evolution of gravitational waves can
be modified in high energy epochs qualitatively.
The wave number is chosen 
$k = a(\tau_{\rm ini}) H(\tau_{\rm ini})$, which corresponds 
to the mode just crossing the cosmic horizon at 
$\tau=\tau_{\rm ini}$.
In Figs.~\ref{fig5.eps} and \ref{fig6.eps}, the scale factors of
our examples are taken to be $a(\tau_{\rm ini})=1.58\times 10^{-17}$ 
and $a(\tau_{\rm ini})=1.58\times 10^{-16}$, respectively.
The expansion of the universe is completely dominated by radiation in
such early epochs. We only consider ordinary radiation (CMB) and three
species of massless neutrinos for simplicity neglecting other
matter components, and set the curvature scale of AdS5 bulk to be $1$
mm.
In this cosmology, the transition from the $\rho^2$ dominated
universe to the standard radiation dominated one occurs at 
$a\approx 4 \times 10^{-17}$.


The evolution of gravitational waves in brane world  differs
that of standard cosmology mainly in two points.
First, the amplitude of gravitational waves in brane world models becomes
smaller than that in four dimensional standard models.
This is due to the fact that
decelerating brane motion excites the modes propagating into the
bulk \cite{Hiramatsu:2003iz}.
Second, the effective frequency $\omega$ of gravitational waves measured
on the brane 
becomes larger due to their momentum not only along the three
dimensional space ($k$ in Eq.(\ref{gweq})) but also to the bulk direction.
These effects becomes remarkable in the high energy epoch ($l^2H^2\gg1$).


\vspace*{1pc}

\section{Summary and Discussion}
\label{sec:summary-discussion}


In this paper, we have carefully investigated the causal structure of
the flat FRW model of RS type II brane world and proposed the
single null coordinate system to solve the cosmological evolution of
gravitational waves.
We have explicitly seen that in this null coordinate system, we do not
have to care the singularity in Eq.~(\ref{eq:GN-GW-eq}).
Further, it is not necessary to introduce any artificial boundaries
to impose some boundary conditions at the bulk infinity.
Thus, we have shown that the problems of the singularity in
Eq.~(\ref{eq:GN-GW-eq}) and the boundary conditions for gravitational
waves at the bulk infinity are resolved if we simply choose an
appropriate initial conditions for gravitational waves.


The initial condition for gravitational waves which we considered to
obtain the numerical examples (Fig.\ref{fig5.eps} and 
\ref{fig6.eps}) in the main text might not be realistic one, since
$h=const.$ is not an exact solution to Eq.(\ref{gweq}) but a mere
approximated solution in the long wavelength limit $k \to 0$. 
Since the aim of this paper is to propose a numerical procedure to
solve the evolution of cosmological gravitational waves, the details
and quantitative studies of numerical results and the problem on
realistic initial conditions of stochastic gravitational waves are
beyond the current scope of this paper and they will be investigated
in our forthcoming paper \cite{ichiki:2004}.
However, it is interesting to discuss the evolution of gravitational
waves with an appropriate initial conditions and clarify the final
spectrum of stochastic gravitational waves resulting from various
creation scenarios of the brane world.


The initial conditions for gravitational waves in brane world
cosmologies crucially depend on the creation scenario of the FRW
brane.
Many kinds of cosmological scenarios have been proposed so
far~\cite{Khoury:2001wf}.
Among them, there are some scenarios where the FRW brane is created
after the inflationary phase.
If we adopt these brane inflationary scenarios, it might be natural to
consider that the initial spectrum of gravitational waves is
determined in this inflationary phase.
The spectrum of gravitational waves under
the deSitter evolution of the brane
is discussed by several
authors~\cite{Langlois:2000ns,Gorbunov:2001ge,Kobayashi:2003cn}.
It was pointed out that gravitational
waves decay away except but the zero-mode, and it approaches to a
constant amplitude in the inflationary phase~\cite{Langlois:2000ns}.
These discussions are based on the GN coordinate system.
Since GN coordinate system does not cover the entire bulk space, these
discussions seem inappropriate to specify the initial spectrum of
gravitational waves.
However, it was also shown that the vacuum defined on the deSitter
slicing asymptotically approaches to the vacuum defined in terms of
the Poincare coordinates on AdS$_{5}$~\cite{Gorbunov:2001ge}.
This will imply that the vacuum state on the static AdS$_{5}$ frame is
appropriate as the bulk initial spectrum of gravitational waves when
we consider these inflationary scenarios.


If we do not adopt the inflationary scenarios, we have to specify
the initial spectrum according to the other creation scenario of the
FRW brane. 
However, in any case, once given the initial configuration on a null
hypersurface, our method can be applied to solve the evolution of
gravitational waves. 
The final spectrum of the stochastic gravitational waves can be
a powerful probe to investigate the existence of extra-dimensions by
comparison with the spectrum in the four-dimensional standard 
cosmology.
We leave this to future works.


Besides the evolution of stochastic gravitational waves, the
procedure developed here will be applicable 
to discuss the evolution of the density perturbations in the brane
world. 
The problem in the choice of the initial spectrum of the
density perturbation will also arise as discussed above and this
initial spectrum will also depend on the creation scenario of the 
FRW brane.
However, according to the causality discussed in this paper, we can
easily expect that the density perturbation is completely determined
by the initial condition on a null hypersurface and the boundary
conditions at the brane. 
Though this expectation should be confirmed by examining the equations
for the density perturbations of the brane world, it is quite
interesting to compare the evolution of the density perturbations in
the brane world scenario with that in the conventional
four-dimensional cosmology. 
We also leave this problem as a future work.


\begin{acknowledgments}
We would like to thank H. Ishihara and T. Tanaka for useful suggestions.
This work has been supported in part by 
the Sasakawa Scientific Research Grant from The Japan Science Society.
We also thank anonymous referees for pointing out our misleading
presentations and improving the quality of our paper.
KN would like to thank all members of Department of Physics of
Hiyoshi Campus in Keio University, all members of Division of
Theoretical Astronomy in NAOJ, and the other colleague for their
continuous encouragement.
\end{acknowledgments}

\appendix
\section{Gaussian normal geodesics}
\label{sec:t_0-derivation}

We briefly review the relation between the Gaussian normal (GN)
coordinates and the flat chart in AdS$_{5}$.
To do this, we first consider the spacelike geodesics normal to the
brane. 
Let us consider the unit normal vector $n^{a}$ ($g_{ab}n^{a}n^{b}=1$)
tangent to these spacelike geodesics. 
The existence of the Killing field $(\partial/\partial t_0)^a$ in the
bulk spacetime ensures that the existence of the constant 
${\cal E} = - g_{ab}n^a (\partial /\partial t_0)^b$ along the geodesics.
Using this constant, the components of the normal vector $n^a$ are given
by 
\begin{eqnarray}
  n^{a} &=:& 
  n^{t_{0}} \left(\frac{\partial}{\partial t_{0}}\right)^{a}
  + n^{r_{0}} \left(\frac{\partial}{\partial r_{0}}\right)^{a}
  \nonumber\\
  &=& \frac{{\cal E}}{f(r_{0})} 
  \left(\frac{\partial}{\partial t_{0}}\right)^{a}
  - \sqrt{f(r_{0}) + {\cal E}^{2}} 
  \left(\frac{\partial}{\partial r_{0}}\right)^{a}.
  \label{tangentvector}
\end{eqnarray}


The explicit orbit of these geodesics is
given by the integration of $dx^\mu=n^\mu dw$.  
For $r_0$ component, we obtain 
\begin{equation}
 2r_0^2+l^2{\cal E}^2=l^2{\cal E}^{2} \cosh\left[-\frac{2}{l}(w-w_0)\right] ~~,
\label{r_0integrated}
\end{equation}
where $w_0$ is the additional constant of integration.
We choose ${\cal E}$ and $w_{0}$ by the conditions: (i) the geodesics are
normal to the brane world volume of $r_0=R(t_0)=a(\tau)$ at $t_0=t_b$ 
(i.e. $n_a \propto \nabla_a\left[r_0-R(t_0)\right]$); 
(ii) $w=0$ on the hypersurface. 
Since $dt_{0}/d\tau$ along the brane is given by the definition of the
cosmic time $\tau$:
\begin{eqnarray}
-d\tau^2&=&-f(r_0)dt_0^2+f^{-1}(r_0)dr_{0}^2 \nonumber \\
 &=&-f(r_0)dt_0^2+f^{-1}(r_0){\dot a}^2 d\tau^2 ~~,
\label{dtau_dt0}
\end{eqnarray}
the above two conditions lead 
\begin{eqnarray}
 {\cal E}(t_b)&=&-\dot a(\tau) ~~, \nonumber \\
 w_0(t_b)&=&\frac{l}{2}\cosh^{-1}\left[\frac{2a^2+l^2{\cal
 E}^2(t_b)}{l^2{\cal E}^{2}(t_b)}\right] ~~.
  \label{gncond}
\end{eqnarray}
Thus, substituting eqs.(\ref{gncond}) into eq.(\ref{r_0integrated}), we
obtain eq. (\ref{r_0oftauw}).
On the other hand, the integration of the $t_{0}$ component of
$dx^\mu=n^\mu dw$, i.e.,
\begin{equation}
 \frac{dt_0}{dw}=n^{t_0}=-\frac{l^2\dot a(\tau)}{\varphi(\tau,w)a^2(\tau)}~~.
\end{equation}
leads Eq.~(\ref{t0astauw}).


Finally, we summarize some useful relations. 
The fact that $\tau$ is constant along these spacelike geodesics leads
\begin{eqnarray}
 \left(\frac{\partial t_0}{\partial w}\right)_{\tau}&=&n^{t_0}=-\frac{\dot a}{f(r_0)} ~~,\label{A6}\\
 \left(\frac{\partial r_0}{\partial w}\right)_{\tau}&=&n^{r_0}=-\sqrt{f(r_0)+{\dot a^2}}~~.
\end{eqnarray}
Also, from the definition that $dw=g_{\mu \nu}n^\mu dx^\nu$, we obtain 
\begin{eqnarray}
\left(\frac{\partial w}{\partial t_0}\right)_{r_0}&=&\dot a(\tau) ~~,\\
\left(\frac{\partial w}{\partial r_0}\right)_{t_0}&=&-\frac{\sqrt{f(r_0)+{\dot a^2(\tau)}}}{f(r_0)}~~.
\label{A9}
\end{eqnarray}
These relations are used in the main text (see Sec.\ref{sec:eq-of-GW}).



%

\begin{thebibliography}{0}

\bibitem{Randall:1999vf} 
L.~Randall and R.~Sundrum,
Phys.\ Rev.\ Lett.\  {\bf 83}, 4690 (1999) 
[arXiv:hep-th/9906064].

\bibitem{Binetruy:1999hy} 
P.~Binetruy, C.~Deffayet, U.~Ellwanger and D.~Langlois,
Phys.\ Lett.\ B {\bf 477}, 285 (2000) 
[arXiv:hep-th/9910219];
P.~Kraus,
JHEP {\bf 9912}, 011 (1999)
[arXiv:hep-th/9910149];
S.~Mukohyama,
Phys.\ Lett.\ B {\bf 473}, 241 (2000)
[arXiv:hep-th/9911165].


\bibitem{Kodama:2000fa}
H.~Kodama, A.~Ishibashi and O.~Seto,
Phys.\ Rev.\ D {\bf 62}, 064022 (2000)


\bibitem{Langlois:2000iu}
D.~Langlois, R.~Maartens, M.~Sasaki and D.~Wands,
Phys.\ Rev.\ D {\bf 63}, 084009 (2001)
[arXiv:hep-th/0012044];
%
%
K.~Koyama and J.~Soda,
Phys.\ Rev.\ D {\bf 65}, 023514 (2002)
[arXiv:hep-th/0108003].


\bibitem{Hiramatsu:2003iz}
T.~Hiramatsu, K.~Koyama and A.~Taruya,
Phys.\ Lett.\ B {\bf 578}, 269 (2004)
[arXiv:hep-th/0308072].
%
\bibitem{Easther:2003re}
R.~Easther, D.~Langlois, R.~Maartens and D.~Wands,
JCAP {\bf 0310}, 014 (2003)
[arXiv:hep-th/0308078].

\bibitem{Ishihara:2001qe}
H.~Ishihara,
Phys.\ Rev.\ D {\bf 66}, 023513 (2002)
[arXiv:gr-qc/0107085]
;
H.~Ishihara and I.~Tanaka, Proceedings of the 5th RESCEU, p59 (2001)
.


\bibitem{Nakamura:2002bz}
K.~Nakamura,
Phys.\ Rev.\ D {\bf 66}, 084005 (2002)
[arXiv:gr-qc/0205031].

\bibitem{Ichiki:2002eh}
K.~Ichiki, M.~Yahiro, T.~Kajino, M.~Orito and G.~J.~Mathews,
Phys.\ Rev.\ D {\bf 66}, 043521 (2002)
[arXiv:astro-ph/0203272].

\bibitem{Spergel:2003cb}
D.~N.~Spergel {\it et al.}, to be appear in ApJ,
arXiv:astro-ph/0302209.

\bibitem{Mukohyama:1999wi}
S.~Mukohyama, T.~Shiromizu and K.~i.~Maeda,
Phys.\ Rev.\ D {\bf 62}, 024028 (2000)
[Erratum-ibid.\ D {\bf 63}, 029901 (2001)]
[arXiv:hep-th/9912287].


\bibitem{ichiki:2004}
K. Ichiki and K. Nakamura,
[arXiv:astro-ph/0406606].

\bibitem{Khoury:2001wf}
J.~Khoury, B.~A.~Ovrut, P.~J.~Steinhardt and N.~Turok,
Phys.\ Rev.\ D {\bf 64}, 123522 (2001)
[arXiv:hep-th/0103239];
J.~J.~Blanco-Pillado and M.~Bucher,
Phys.\ Rev.\ D {\bf 65}, 083517 (2002)
[arXiv:hep-th/0111089];
J.~Garriga and T.~Tanaka,
Phys.\ Rev.\ D {\bf 65}, 103506 (2002)
[arXiv:hep-th/0112028];
U.~Gen, A.~Ishibashi and T.~Tanaka,
Phys.\ Rev.\ D {\bf 66}, 023519 (2002)
[arXiv:hep-th/0110286]
.


\bibitem{Langlois:2000ns}
D.~Langlois, R.~Maartens and D.~Wands,
Phys.\ Lett.\ B {\bf 489}, 259 (2000)
[arXiv:hep-th/0006007].

\bibitem{Gorbunov:2001ge}
D.~S.~Gorbunov, V.~A.~Rubakov and S.~M.~Sibiryakov,
JHEP {\bf 0110}, 015 (2001)
[arXiv:hep-th/0108017].

\bibitem{Kobayashi:2003cn}
T.~Kobayashi, H.~Kudoh and T.~Tanaka,
Phys.\ Rev.\ D {\bf 68}, 044025 (2003)
[arXiv:gr-qc/0305006].

\end{thebibliography}
\end{document}